\documentclass[preprint]{aastex}

\usepackage{epsf}

\shorttitle{High-Redshift Galaxies in CDM}
\shortauthors{Weinberg, Katz, \& Hernquist}

\newcommand{\lya}{Ly$\alpha$\ }
\newcommand{\kms}{{\rm km}\;{\rm s}^{-1}}
\newcommand{\hubunits}{\kms\;{\rm Mpc}^{-1}}

\newcommand{\hmpc}{h^{-1}\;{\rm Mpc}}
\newcommand{\hkpc}{h^{-1}\;{\rm kpc}}
\newcommand{\msun}{M_\odot}
\newcommand{\msunyr}{M_\odot\;{\rm yr}^{-1}}
\newcommand{\K}{{\rm K}}
\newcommand{\cm}{{\rm cm}}

\newcommand{\be}{\begin{equation}}
\newcommand{\ee}{\end{equation}}
\newcommand{\asfr}{{\langle {\rm SFR}\rangle}}
\newcommand{\Om}{\Omega_m}
\newcommand{\Omb}{\Omega_b}
\newcommand{\Ol}{\Omega_\Lambda}

\begin{document}

\title{HIGH-REDSHIFT GALAXIES IN COLD DARK MATTER MODELS}

\author{David H. Weinberg\altaffilmark{1},
Lars Hernquist\altaffilmark{2},
and Neal Katz\altaffilmark{3}}
\altaffiltext{1}
{Ohio State University, Department of Astronomy, Columbus, OH 43210,
dhw@astronomy.ohio-state.edu}
\altaffiltext{2}
{Harvard-Smithsonian Center for Astrophysics, Cambridge, MA 02138,
lars@cfa.harvard.edu}
\altaffiltext{3}
{University of Massachusetts, Department of Physics and Astronomy,
Amherst, MA 91003, nsk@kaka.phast.umass.edu}

\begin{abstract}

We use hydrodynamic cosmological simulations to predict the star formation 
properties of high-redshift galaxies ($z=2-6$) in five variants of the
inflationary cold dark matter scenario, paying particular attention to $z=3$,
the redshift of the largest ``Lyman-break galaxy'' (LBG) samples.  Because
we link the star formation timescale to the local gas density, the rate at
which a galaxy forms stars is governed mainly by the rate at which it 
accretes cooled gas from the surrounding medium.  At $z=3$, star formation
in most of the simulated galaxies is steady on $\sim 200$ Myr timescales, and
the instantaneous star formation rate (SFR) is correlated with total stellar
mass.  However, there is enough scatter in this correlation that a sample
selected above a given SFR threshold may contain galaxies with a fairly wide
range of masses.  The redshift history and global density of star formation 
in the simulations depend mainly on the amplitude of mass fluctuations in the 
underlying cosmological model.  The three models whose mass fluctuation 
amplitudes agree with recent analyses of the \lya forest also reproduce the 
observed luminosity function of LBGs reasonably well, though the dynamic range 
of the comparison is small and the theoretical and observational uncertainties 
are large.  The models with higher and lower amplitudes appear to predict too 
much and too little star formation, respectively, though they are not clearly 
ruled out.  The intermediate amplitude models predict 
SFR$\;\sim 30-40\msunyr$ for galaxies with a surface density 
$\sim 1\;{\rm arcmin}^{-2}$ per unit redshift at $z=3$.  They predict much
higher surface densities at lower SFR, and significant numbers of galaxies
with SFR$\;>10\msunyr$ at $z \geq 5$.

\end{abstract}

\keywords{galaxies: formation, dark matter, large-scale
structure of the Universe}

\section{Introduction}
\label{sec:intro}

The discovery and characterization of ``Lyman-break'' galaxies (LBGs) has
opened a new window on the high-redshift universe, revealing a 
population of star-forming galaxies at $z>3$ whose comoving space
density exceeds that of $L_*$ galaxies today
\citep{steidel96,lowenthal97}.
These galaxies can be identified by their unusual colors in 
deep imaging surveys because the intrinsic continuum break
at $\lambda \sim 912$\AA\ and the intergalactic absorption
by the \lya forest at $\lambda < 1216$\AA\ redshift into optical bands.
Spectroscopic follow-up shows that photometry of Lyman-break
objects yields robust approximate redshifts.
{}From an optical imaging survey, one can therefore construct
a sample of high-$z$ galaxies limited primarily by rest-frame
ultraviolet (UV) luminosity, which, in the absence of dust extinction,
is itself determined mainly
by the instantaneous formation rate of massive stars.
Application of this approach to the Hubble Deep Field 
\citep[HDF; ][]{williams96}
and other deep imaging surveys has yielded first attempts
at one of the long-standing goals of observational cosmology, 
determination of the star formation history of the universe 
\citep[e.g.,][]{madau96,madau97,connolly97,steidel99}.

In this paper, we examine the ability of models based on inflation
and cold dark matter (CDM) to account for the observed population
of LBGs, using cosmological simulations that
incorporate gravity, gas dynamics, and star formation.
We consider five variants of the CDM scenario: three $\Om=1$
models, a spatially flat low density model with a cosmological constant,
and an open universe low density model with $\Ol=0$.
The spatial clustering of the high-redshift galaxies in these
simulations was discussed by \citet{katz99};
here we focus on the masses and star formation properties of
these galaxies.

Numerical simulations play two overlapping but distinct
roles in cosmological studies.  First, they provide
quantitative predictions that can be compared to observations
in order to test the underlying cosmological models.
Second, they provide greater understanding of the observational
phenomena themselves, by showing how observable structures might
arise and evolve in a given cosmological scenario.
In this paper we will emphasize the second of these roles, mainly
because the numerical limitations of the simulations and our limited
knowledge of the physics of star formation contribute uncertainties
that are comparable to the differences between cosmological models.
The examination of different cosmologies is still a useful exercise,
however, because it shows how cosmological parameters and properties
of primordial mass fluctuations affect the properties of the 
high-redshift galaxy population when other physical and numerical
parameters are held fixed.

Hydrodynamic simulations complement the main alternative approach
to the theoretical study of high-redshift galaxies, based on
semi-analytic models of galaxy formation 
\citep[e.g.,][]{baugh98,kauffmann99,somerville00}.
Semi-analytic models have the advantages of simplicity, flexibility, and speed. 
The price is a substantial number of approximations and tunable parameters;
the values of some parameters are fixed by matching selected observations,
leaving other observables as predictions of the model.
Semi-analytic models incorporate simplified descriptions of gravitational 
collapse, mergers, and cooling of gas within dark halos.
The strength of numerical simulations is their more realistic
treatment of these processes.  The only free parameters (apart from
the physical parameters of the cosmological model being studied)
are those related to the treatment of star formation and feedback.
Given these parameters, simulations provide straightforward, untunable 
predictions.  However, the simulation approach must contend with
the numerical uncertainties caused by finite volume and finite resolution,
and computational expense makes it a slow way to explore parameter space.
Over the next few years, interactions between the numerical and
semi-analytic approaches should strengthen both.  Here we mainly 
present the numerical results on their own terms, with a brief
comparison to interpretations based on semi-analytic models in
\S\ref{sec:disc}. 

We describe our numerical methods, treatment of star formation,
and choice of cosmological model in \S\ref{sec:sims}.  In \S\ref{sec:gals}
we present results for the LCDM model (CDM with a cosmological constant) 
at $z=3$, the redshift best probed by recent Lyman-break galaxy surveys.
In \S\ref{sec:csf} we broaden our scope, examining predictions of five
different CDM models for the population of star-forming galaxies 
from $z=6$ to $z=2$.  We discuss our results and prospects for
future progress in \S\ref{sec:disc}.

\section{Simulations}
\label{sec:sims}

\subsection{Numerical Parameters and Star Formation}
\label{sec:numerics}

All of our simulations use TreeSPH \citep[][hereafter KWH]{hernquist89,katz96},
a code that combines smoothed particle hydrodynamics
\citep[SPH; see][]{lucy77,gingold77,monaghan92}
with a hierarchical tree algorithm \citep{barnes86} for computing 
gravitational forces.  The method and illustrative cosmological
applications are described in detail by KWH, so here we just 
specify the simulation parameters and recap
the points that are most important to the present investigation.

Each of our simulations uses $64^3$ dark matter and $64^3$ SPH particles
to model a triply periodic volume $11.111\hmpc$ comoving Mpc on a side,
where $h \equiv  H_0/(100\;\hubunits)$.  
The simulations are evolved to $z=2$. 
For the three critical density ($\Om=1$) cosmological models,
the dark matter particle mass is $2.76\times 10^9 \msun$ and
the SPH particle mass is $1.45\times 10^8\msun$.
For the two low density ($\Om=0.4$) models,
the dark matter particle mass is $8.27 \times 10^8 \msun$ and
the SPH particle mass is $6.71\times 10^7\msun$.
Gravitational forces are softened using a cubic spline kernel
with a softening length $\epsilon = 5\hkpc$, equivalent to
$\epsilon \approx 3.5\hkpc$ for a Plummer softening law.
The gravitational softening length is held fixed in comoving units,
i.e., $\epsilon=1.25\;h^{-1}$ physical kpc at $z=3$.
Particles have individual time steps that satisfy the conditions
$\Delta t < 0.4{\rm min}(\epsilon/|{\bf v}|,\sqrt{\epsilon/|{\bf a}|})$,
where ${\bf v}$ is the peculiar velocity and ${\bf a}$ is the acceleration.
SPH particle time steps are also required to satisfy the Courant condition
(see KWH).  The maximum time step for any particle is
$\Delta t_d = H_0^{-1}/6000$.

Radiative cooling is computed assuming primordial composition gas
with helium abundance $Y=0.24$ by mass.  All of the simulations incorporate
a photoionizing UV background with the spectral shape and redshift
history computed by \citet{haardt96}, but with intensity reduced
by a factor of two in order to approximately match the mean opacity
of the \lya forest given our assumed baryon density \citep{croft97}.
In practice, the photoionizing background has negligible effect on
the Lyman-break galaxy population, at least in the mass range that
our simulations can resolve \citep{weinberg97a}.

The gas that resides in collapsed dark matter halos exhibits a two-phase
structure: hot gas at roughly the halo virial temperature with a density
profile similar to that of the dark matter (but exhibiting a core
at small radii), and radiatively cooled gas
with $T \sim 10^4\;\K$ at much higher overdensity.
In simulations that do not incorporate star formation, the
clumps of radiatively cooled gas have masses and sizes comparable
to the luminous regions of observed galaxies.
Our star formation algorithm is essentially a prescription for turning
this dense, cold gas into collisionless stars, returning energy
from supernova feedback to the surrounding medium.
We provide a brief synopsis of this algorithm here and refer the
reader to KWH for details.

An SPH gas particle is ``eligible'' to form stars if it
is Jeans unstable, resides in a region of converging flow,
has an overdensity $\rho_g / \overline{\rho_g} > 55.7$
(corresponding to the virial boundary of a singular isothermal sphere
in the spherical collapse
model), and has a hydrogen number density exceeding $0.1\;\cm^{-3}$
(physical units).
In practice, it is the physical density threshold that matters ---
gas with this density almost always satisfies the other criteria,
except at very high redshift, where the overdensity threshold
ensures that star formation does not occur in uncollapsed regions
simply because the cosmic mean density is high.
Once a gas particle is {\it eligible} to form stars,
its star formation {\it rate} is given by
\begin{equation}
{d\rho_\star \over dt} = - {d\rho_g \over dt} =
		       {c_\star\epsilon_\star\rho_g \over t_g},
\label{eqn:sfrate}
\end{equation}
or
\begin{equation}
{d\ln\rho_g \over dt} = -{c_\star\epsilon_\star \over t_g},
\label{eqn:lnsfrate}
\end{equation}
where $c_\star$ is a
dimensionless star formation rate parameter,
$\epsilon_\star$ is the fraction of the particle's gas mass that
will be converted to stellar mass in a single simulation timestep,
and the gas flow timescale $t_g$ is the maximum of the local gas dynamical 
time, $t_{\rm dyn}=(4\pi G\rho_g)^{-1/2}$, and the local cooling time.
Each SPH particle has both a gas mass and a stellar mass (initially zero);
the total gas$+$stellar mass contributes to gravitational forces,
but only the gas mass is used in computing the SPH properties and forces.
The probability $p$ that an eligible SPH particle undergoes a star
formation event in an integration timestep of duration $\Delta t$ is
\begin{equation}
p = 1-\exp\left({-c_* \Delta t \over t_g}\right),
\end{equation}
and if the particle does undergo such an event then $\epsilon_\star=1/3$
of its remaining gas mass is converted into stars during that step.
In the limit (nearly always satisfied in the simulations) that
$c_*\Delta t / t_g \ll 1$, this algorithm yields the average
star formation rate given by 
equation~(\ref{eqn:sfrate}).\footnote{In KWH, the description of
the algorithm is accurate but their equations (44) and (45), which correspond
to equations~(\ref{eqn:sfrate}) and~(\ref{eqn:lnsfrate}), are missing
the factor of $\epsilon_\star$.}
Once a particle's gas mass falls below 5\% of its original mass,
it is converted into a collisionless, pure star particle, affected only
by gravity, and its residual gas mass is redistributed to its SPH neighbors.

When an SPH particle undergoes star formation, recycled gas and
supernova feedback energy are distributed to the particle and its
neighbors, assuming a Miller-Scalo (\citeyear{miller79}) 
initial mass function truncated
at $0.1\msun$ and $100\msun$ and $10^{51}$ ergs per supernova.
This feedback energy is usually radiated away because it is released
into a dense, gas rich medium with a short cooling time.
Feedback therefore has only a modest impact in our simulations,
and this is the physically appropriate result if the proto-galactic
interstellar medium is fairly smooth and as dense as our simulations imply.
It is possible that strong inhomogeneities in the interstellar medium
(on scales well below our resolution limits) allow feedback to have
a stronger effect in real proto-galaxies, and explicit modeling of
this possibility is an important direction for future investigation.
The scenario that we investigate here is a physically plausible
limiting case.

In all of our simulations we set the star formation rate parameter
$c_\star$ to 0.1 and $\epsilon_\star$ to 1/3.  
As shown in KWH, the stellar masses 
of the simulated galaxies are insensitive to the value of $c_\star$.
In the KWH tests, an order of magnitude increase to $c_\star=1.0$ 
changes the total stellar mass in the box at $z=2$ by only 15\%,
and the effect of a higher $c_\star$ is actually to decrease the
stellar mass because star formation occurs in lower density gas
where supernova feedback can have a stronger effect.
Indeed, one obtains nearly the same galaxy population in simulations
that do not include star formation at all, except that in this case
the ``galaxies'' are the clumps of cold, dense gas instead
of the clumps of cold, dense gas and stars (see KWH, figure 5).
In our simulations, the rate at which a galaxy forms stars is
governed mainly by the rate at which gas condenses from the hot halo
into the cold clump; the regulation implied by equation~(\ref{eqn:sfrate}) 
ensures that the gas condensation rate and the star formation rate
cannot get too far out of step.
The link between star formation rate and gas density is physically
motivated, since denser gas is more gravitationally unstable and
more easily able to radiate its energy.  In the case where the
cooling timescale is short and $t_g=t_{\rm dyn}$, equation~(\ref{eqn:sfrate})
implies $\dot\rho_* \propto \rho_g^{3/2}$, similar to the
Schmidt-law $\dot\rho_* \propto \Sigma_g^{3/2}$ observed to hold
over a large dynamic range in a wide variety of local galaxies
\citep{schmidt59,kennicutt98}.

\subsection{Cosmological Models}

We consider five different cosmological models, all of which assume 
Gaussian primordial fluctuations and a universe dominated by cold,
collisionless dark matter.  In all cases we adopt a baryon density
parameter $\Omb = 0.0125 h^{-2}$ based on \citet{walker91},
though a higher $\Omb$ is suggested by recent analyses of 
the \lya forest opacity \citep{rauch97,weinberg97b} and the
deuterium abundance in high-redshift Lyman limit systems
(\citealt{burles97}, \citeyear{burles98}).
The model we refer to as ``standard'' CDM (SCDM) assumes $\Om=1$,
$h=0.5$, and an rms linear theory fluctuation in $8\hmpc$
spheres of $\sigma_8=0.7$.  For this model we use the parameterization of the
CDM power spectrum given by \citet{bardeen86}.
The $\sigma_8=0.7$ normalization is roughly consistent with the
observed abundance of rich galaxy clusters \citep{white93}, but
the SCDM model does not reproduce the amplitude of cosmic microwave
background anisotropies observed by the COBE-DMR experiment
\citep{smoot92,bennett96}.
Our second model, COBE-normalized CDM (CCDM),
is the same as SCDM except that the normalization $\sigma_8=1.2$
is chosen to match the 4-year COBE data (\citealt{gorski96}; see
\citealt{bunn97} for a discussion of the CDM normalization).
With this value of $\sigma_8$ and $\Om=1$, the CCDM 
model produces galaxy clusters that are too massive to be consistent
with observations.

One way to reconcile the COBE-DMR anisotropies and the observed
cluster abundance within the context of $\Om=1$ CDM models
is to assume that inflation generates a primeval power spectrum
that is ``tilted'' instead of scale-invariant, $P(k) \propto k^n$
with $n<1$.  For our tilted CDM (TCDM) model, we adopt $n=0.80$
and the transfer function given by equation (D28) of \citet{hu96},
which treats baryon damping effects more accurately than the original 
\citet{bardeen86} formulation.  We normalize this analytic
fit to the power spectrum to the amplitude $\sigma_8=0.54$ implied by 
COBE normalization, which we compute
using the CMBFAST code of 
Seljak \& Zaldarriaga (\citeyear{seljak96}; \citealt{zaldarriaga98}),
assuming the standard tensor mode contribution to microwave background
anisotropies predicted by power law inflation models.  

Another way to resolve the COBE/cluster conflict is to lower the
value of $\Om$, reducing cluster masses for a given $\sigma_8$.
We consider two different low-$\Om$ CDM models, one (LCDM)
with a spatially flat universe and a cosmological constant
$\Ol=1-\Om$ and one (OCDM) with an open universe
and $\Ol=0$.  For LCDM we adopt $\Om=0.4$, $h=0.65$, 
and a primeval spectral index $n=0.93$.  With the tensor mode
contribution, CMBFAST implies a normalization
$\sigma_8=0.8$, which provides a good match to the cluster
abundances for $\Om=0.4$ \citep{white93}.
We again use the \citet{hu96} formulation of the
transfer function.  For OCDM, we adopt $\Om=0.4$, $h=0.65$,
$n=1.0$, and a 2-year COBE-DMR normalization
$\sigma_8=0.75$ \citep{ratra97}.  Cluster masses in this model
are lower than those in TCDM or LCDM, but they are consistent with
current observations given their uncertainties \citep{cole97}.
For OCDM we use the transfer function of \citet{efstathiou92}
with $\Gamma=0.234$; the \citet{hu96} formulation
is more accurate, but we were unaware of it at the time we ran
the OCDM simulation.  In practice, the differences between different
analytic or numerical formulations of transfer functions are of
the same magnitude as the changes caused by slight shifts in the
adopted values of $h$, $\Omb$, or $\Om$.

\begin{figure*}
\centerline{
\epsfxsize=4.5truein
\epsfbox[90 415 470 725]{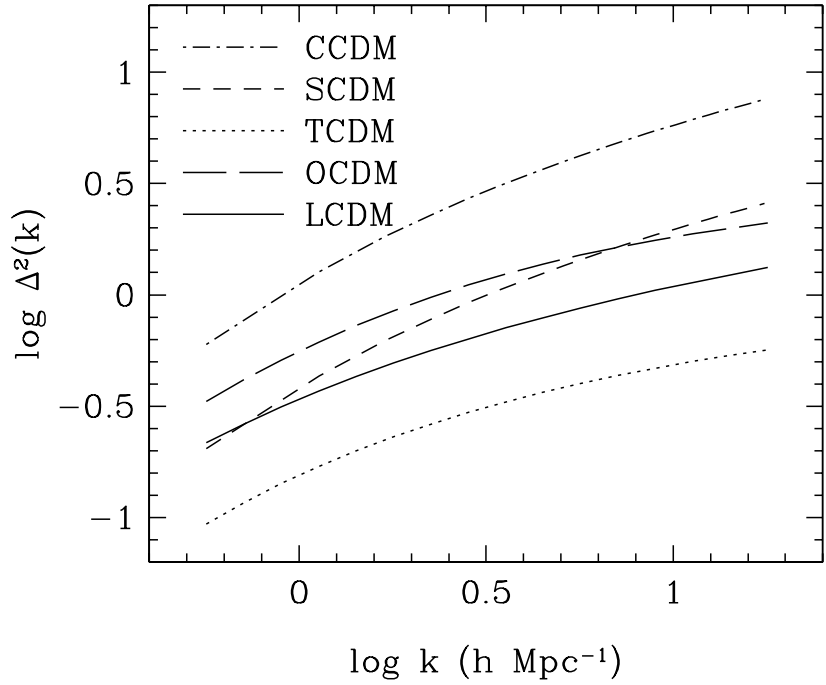}
}
\caption{
\label{fig:pk}
Amplitude of linear mass fluctuations in the different cosmological
models at $z=3$.  The quantity $\Delta^2(k) \equiv 4\pi k^3 P(k)$
is the contribution to the variance of fluctuations per unit interval
of $\ln k$.  The wavenumber $k$ is in comoving $h\;{\rm Mpc}^{-1}$.  
The range shown is from the fundamental mode of the simulation cube,
$k_f = 2\pi / (11.111\hmpc)$, to the Nyquist frequency of the initial
particle grid, $k_N=32k_f$.
}
\end{figure*}

Parameters of the five cosmological models are listed in Table~\ref{tbl-1}.
Figure~\ref{fig:pk} shows the linear theory power spectra of the five
models, over the range of scales represented in the initial conditions
of our simulations.  Instead of $P(k)$ itself, we plot
$\Delta^2(k) \equiv 4\pi k^3 P(k)$, which (with our Fourier transform
convention) is the contribution to the variance of linear mass 
fluctuations per unit interval of ln$k$ \citep[see][]{peacock94}.
The differences in fluctuation amplitude among the three $\Om=1$
models are easy to see, as is the difference in $P(k)$ shape between
the CCDM/SCDM models and the tilted (TCDM) and low density (OCDM/LCDM)
models.  Although LCDM has a slightly higher normalization than 
OCDM at $z=0$ ($\sigma_8=0.80$ vs.\ $\sigma_8=0.75$), OCDM has higher
amplitude fluctuations at $z=3$ because of the smaller ratio of 
linear growth factors between $z=3$ and $z=0$ in an open universe.

\section{Lyman-Break Galaxies in the LCDM Model}
\label{sec:gals}

\begin{figure*}
\centerline{
\epsfxsize=6.0truein
\epsfbox[60 230 550 720]{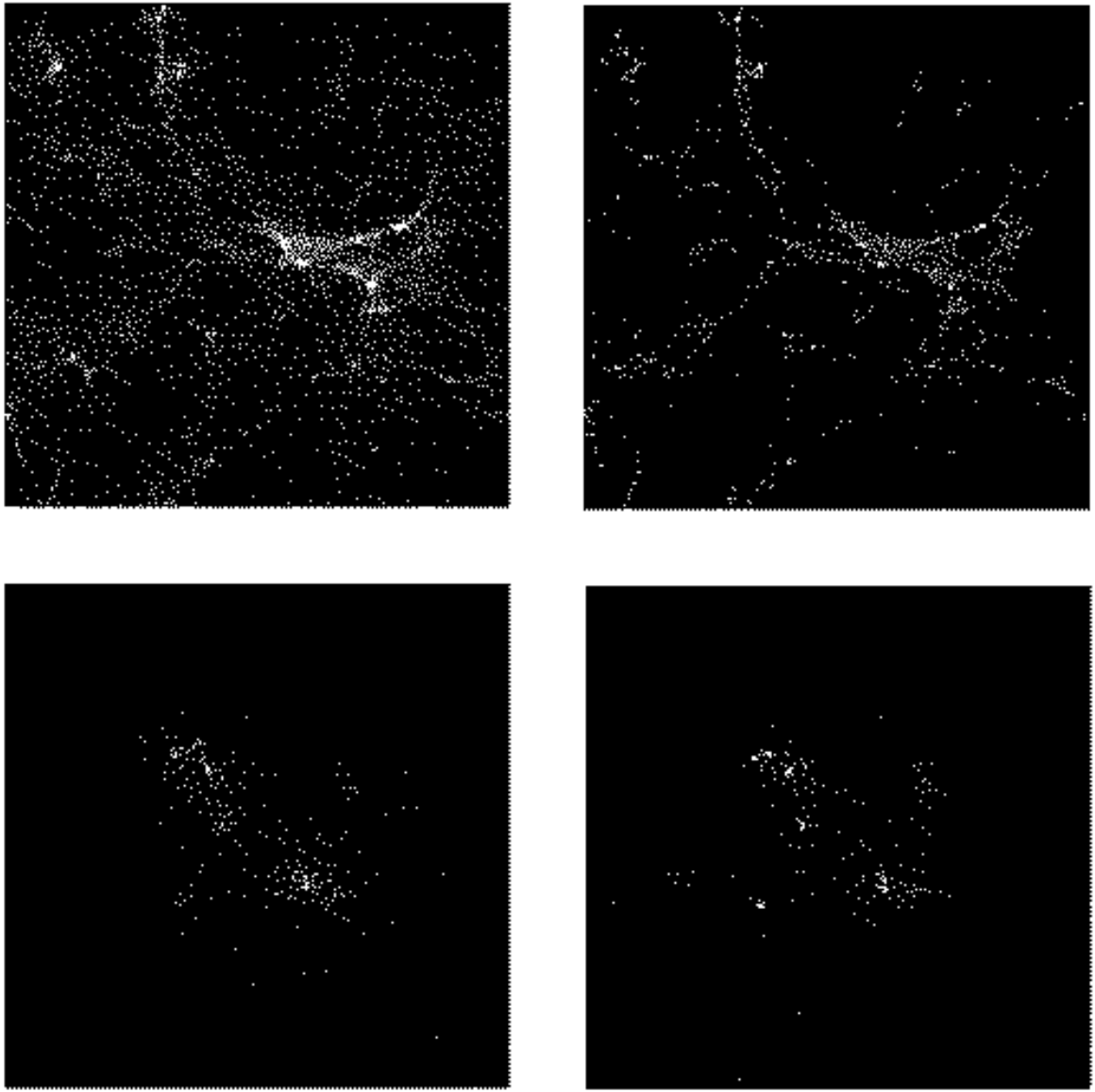}
}
\caption{
\label{fig:color}
Gas and galaxies in the LCDM simulation at $z=3$.
Upper panels show the gas distribution in the full $11.111\hmpc$
simulation cube (comoving size),
color-coded by $\log T$ (left) and by $\log\rho_g$ (right).
The lower left panel shows the gas temperature in a cubical sub-volume 
$0.09\hmpc$ (comoving) on a side.  
The lower right panel shows the dark matter (blue),
gas (red) and star (yellow) particles in the same sub-volume.
The temperature scale in the left hand panels runs from several thousand K
(blue) to $10^4\;\K$ (magenta), $10^5\;\K$ (orange/yellow), and
$\geq 10^6\;\K$ (white).  {\bf Color GIF file provided separately
on astro-ph.}
}
\end{figure*}

We will focus in this Section on the galaxy population of the LCDM
model at $z=3$, then turn to other models and other redshifts in \S 4.
Figure~\ref{fig:color} shows particle distributions at $z=3$ from
the full $11.111\hmpc$ simulation cube (top panels) and a $0.09\hmpc$
subcube containing the richest concentration of galaxies (bottom panels).
The top panels show numerous concentrations of dense, shock heated
gas, with typical temperatures $T \sim 10^6\;\K$ corresponding to 
the virial temperatures of the corresponding dark matter halos.
However, the zoomed view at the lower left reveals the two phase
structure that characterizes collapsed regions of the simulation.
Extremely overdense clumps of $10^4\;\K$ gas, typically a few kpc in size,
reside in a background of gas with $T \ga 10^6\;\K$.  These dense
concentrations of cold gas are, of course, the sites of star formation,
as shown in the lower right panel.  Because the knots of cold gas and
stars stand out so clearly from the background, there is no 
ambiguity in identifying the galaxies in such a simulation; one 
only needs an algorithm that picks out these clumps.
We use the SKID algorithm (Spline-Kernel Interpolated DENMAX; see
KWH and \citealt{gelb94}), as implemented by
Stadel, Katz, \& Quinn\footnote{see 
{\tt http://www-hpcc.astro.washington.edu/tools/SKID/}},
which identifies clumps of gravitationally bound particles associated
with a common density maximum.

In collapsing dark matter halos that contain a small number of particles,
the resolution of the SPH calculation becomes too low for the simulation
to follow the cooling of the gas and subsequent star formation.
Gardner et al.\ (\citeyear{gardner97}, \citeyear{gardner00}) 
find that simulated halos with at
least 60 dark matter particles nearly always contain a cold gas
concentration, while halos with fewer particles often do not.
We can therefore resolve the existence of galaxies in halos with 
mass $M>60m_{\rm dark}$, a quantity that we list in Table~\ref{tbl-1}
as an indication of our dark matter mass resolution.
At $z=3$, the halo circular velocity corresponding to this limiting mass
(plus the associated baryon mass) is
$v_c \sim 140\;\kms$ in the $\Om=1$ models
and $v_c \sim 95\;\kms$ in the low density models (see \citealt{gardner97},
equation 3).  Comparison of LCDM simulations with $64^3$ and $128^3$
SPH particles in the same $11.111\hmpc$ volume 
(Gardner et al., in preparation; Aguirre et al., in preparation) 
suggests that the requirement for
accurate estimation of a galaxy's star formation rate via 
equation~(\ref{eqn:sfrate}) is somewhat more stringent,
corresponding to 60 or more particles in the {\it condensed baryon}
phase (cold gas plus stars).  We list $60m_{\rm SPH}$ as an
approximate baryon mass resolution limit for star formation calculations
in Table~\ref{tbl-1}.

\begin{figure*}
\centerline{
\epsfxsize=4.5truein
\epsfbox[90 415 470 725]{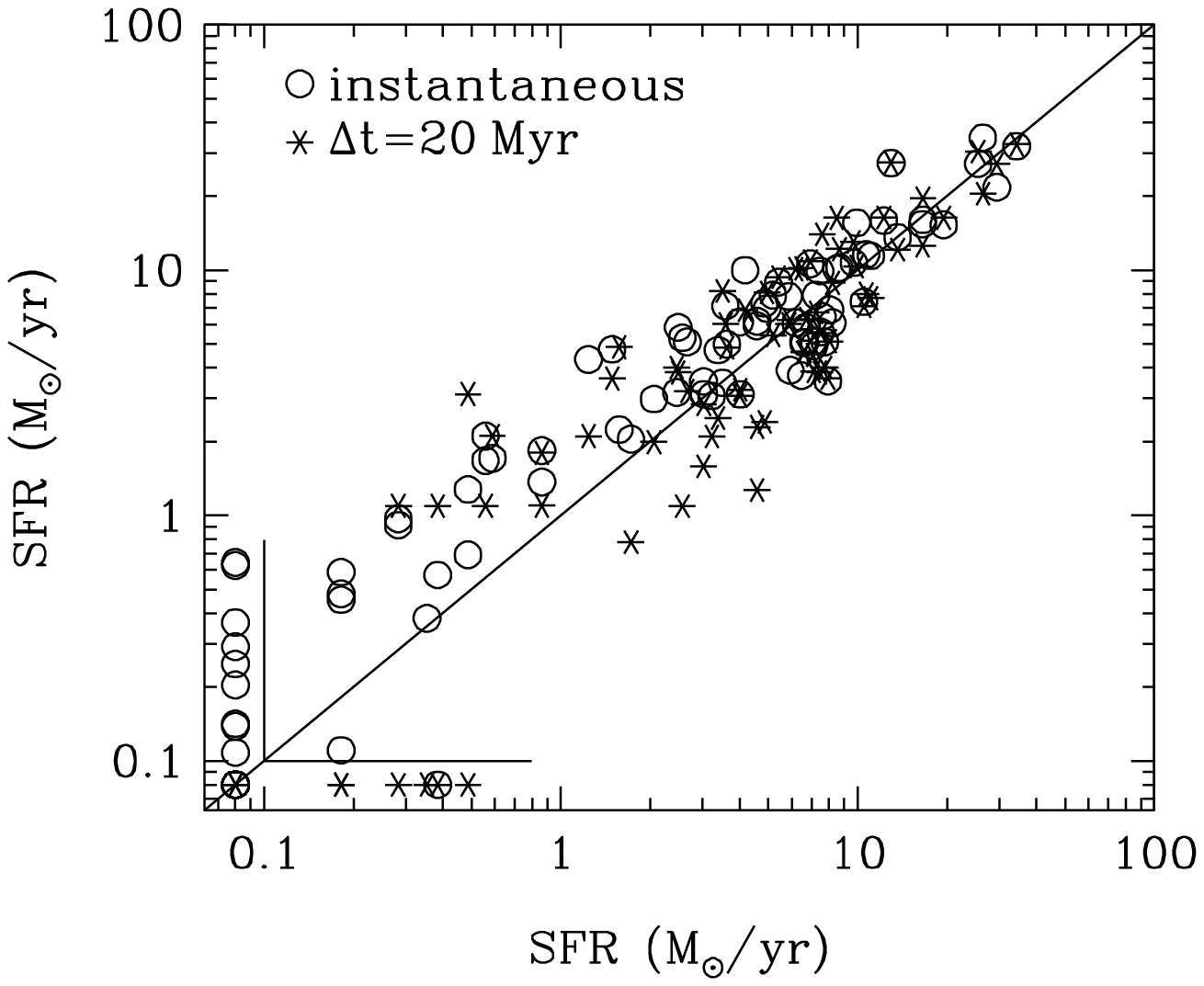}
}
\caption{
\label{fig:sfrComp}
Dependence of star formation rate (SFR) on time-averaging,
for simulated galaxies in the LCDM model at $z=3$.
The $x$-axis is the galaxy star formation rate averaged over
$\Delta t = 200$ Myr.  
Open circles show the instantaneous star formation rate,
estimated by applying the prescription of \S 2 (see eq.~\ref{eqn:sfrate})
to the simulated gas distribution.  Asterisks show the 
star formation rate averaged over $\Delta t = 20$ Myr.
Points along the axis have SFR $<0.1 M_\odot/$yr.
The age of the universe at this redshift is 1.97 Gyr.
}
\end{figure*}

For the galaxies at $z=3$, the asterisks in Figure~\ref{fig:sfrComp}
compare the star formation rate averaged over the previous
20 Myr to the star formation rate averaged over the previous 200 Myr.
The two rates are usually within a factor of two of each other
and are often much closer, indicating that star formation in our
simulated galaxies is fairly steady over intervals of 200 Myr.
Circles in Figure~\ref{fig:sfrComp} show the ``instantaneous'' star
formation rate calculated by applying the prescription of \S 2
to the gas particle distribution.  This is the rate that would be used to
calculate star formation in the simulation's next system timestep
(of duration $\Delta t = 2.5$ Myr).  The tight correlation between
this estimate of the star formation rate
and the rates averaged over longer intervals
demonstrates that the instantaneous estimate is not sensitive
to numerical ``noise'' in particle densities and positions, at
least for systems with star formation rate $> 1 M_\odot/$yr.
We will henceforth use the instantaneous estimate as our
standard measure of the star formation rate (hereafter denoted SFR),
with Figure~\ref{fig:sfrComp} as evidence that our results are
insensitive to this choice.

\begin{figure*}
\centerline{
\epsfxsize=4.5truein
\epsfbox[90 415 470 725]{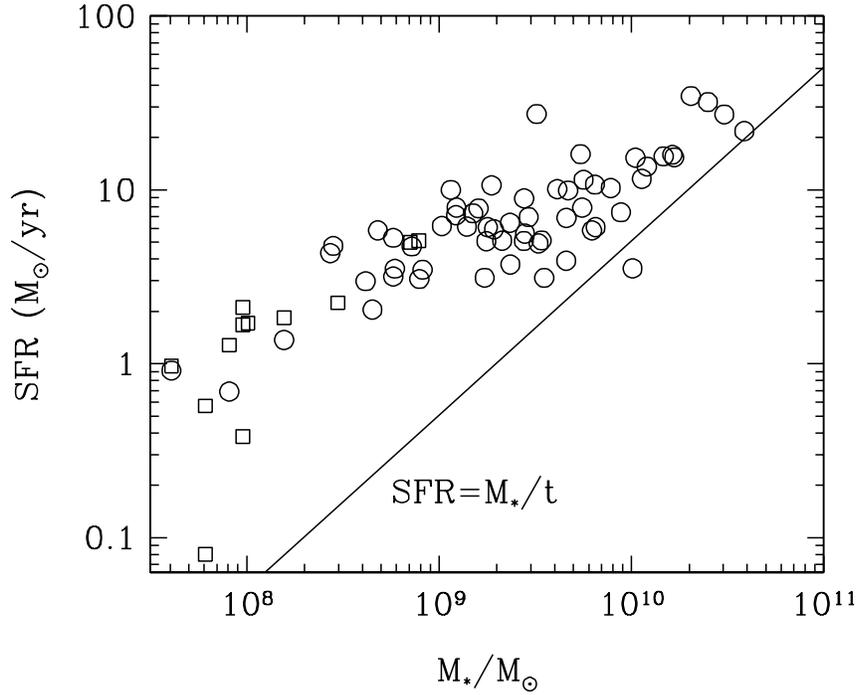}
}
\caption{
\label{fig:sfrVsMstar}
Correlation between instantaneous star formation rate
and stellar mass, for simulated galaxies in the LCDM model at $z=3$.
The solid line shows the relation SFR=$M_*/t$ that would apply for
galaxies forming stars at a constant rate over the age of the
universe $t=1.97$ Gyr.  
Squares represent galaxies with baryonic mass $M<60m_{\rm SPH}$,
for which the SFR is likely to be underestimated because of limited
numerical resolution.
}
\end{figure*}

Figure~\ref{fig:sfrVsMstar} plots galaxies' instantaneous
star formation rates against their stellar masses.
The correlation between SFR and $M_*$ is reasonably strong,
but there is enough scatter that a sample selected by a
threshold in SFR (above a horizontal line in Figure~\ref{fig:sfrVsMstar})
excludes some fairly massive galaxies and includes others that
are substantially further down the stellar mass function.
Nearly all galaxies are forming stars at a rate faster than
the rate $\asfr = M_*/t$ that would build
them up steadily over the age of the universe; this result
is not surprising, since the galaxies do not start to form stars
at $t=0$.  The ratio SFR$/\asfr$ is substantially higher for
low mass galaxies than for high mass galaxies.  Since this
ratio is correlated with the overall shape of a galaxy's
spectral energy distribution, Figure~\ref{fig:sfrVsMstar} implies
that less massive galaxies should be bluer than more massive galaxies.
This trend could be caused partly by our limited numerical 
resolution, since the simulated galaxies do not form stars at the 
correct efficiency until they are sufficiently massive.
However, the trend appears to be present even in the fairly
well resolved systems.

\begin{figure*}
\centerline{
\epsfxsize=4.5truein
\epsfbox[90 415 470 725]{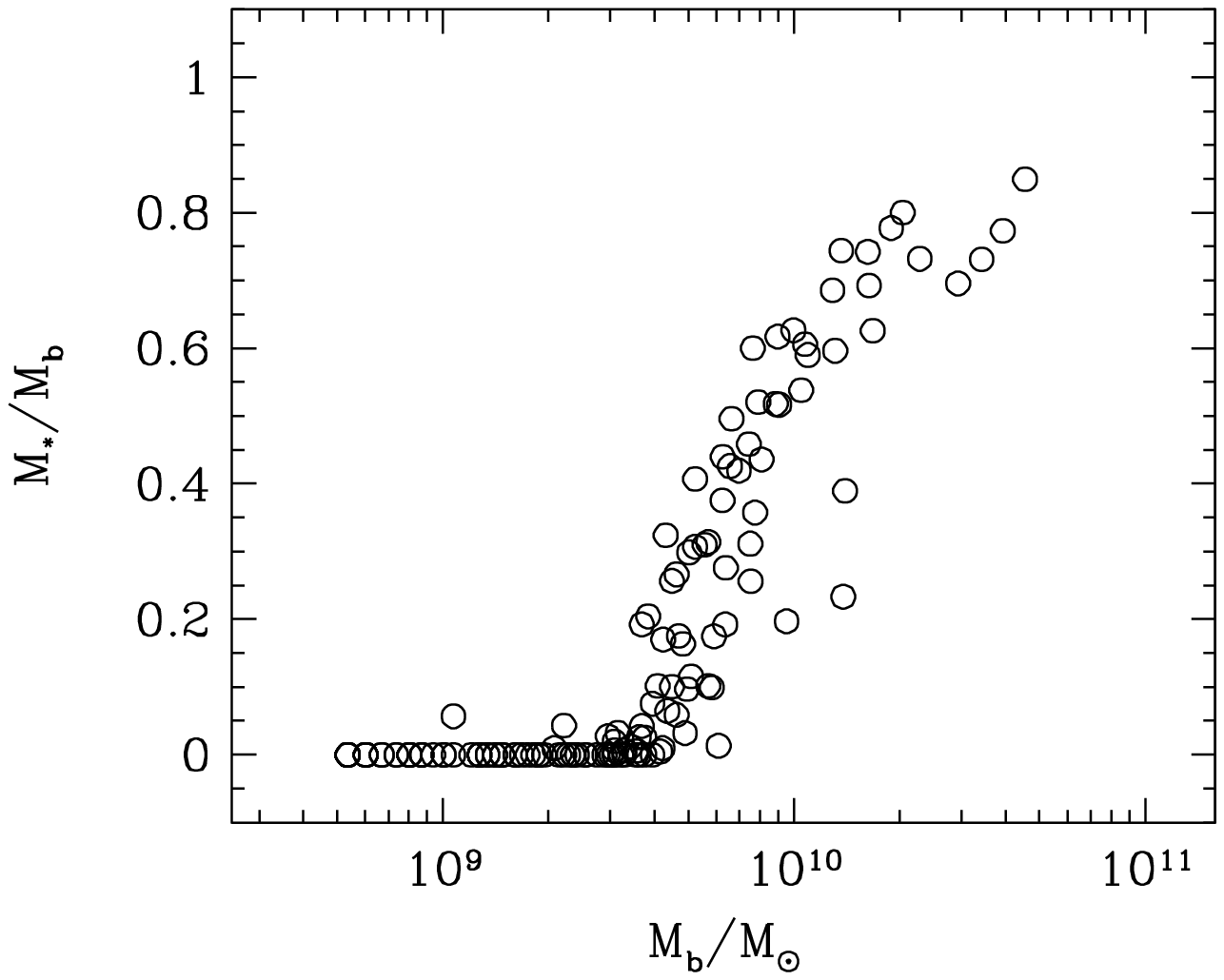}
}
\caption{
\label{fig:starFrac}
Stellar mass fraction as a function of total baryonic mass, for
simulated galaxies in the LCDM model at $z=3$.  $M_*$ is the stellar
mass, and $M_b$ is the total mass of stars and cold, dense gas.
Our estimated minimum mass for correct calculation of the star 
formation rate is $M_b=60m_{\rm SPH}=4\times 10^9 M_\odot$.
However, the trend of stellar mass fraction with total mass may
be exaggerated by numerical resolution effects even above this
threshold, since galaxies near the threshold may have had lower
masses, and underestimated star formation rates, at earlier times.
}
\end{figure*}

Figure~\ref{fig:starFrac} plots the ratio of stellar mass to baryon
mass (stars plus cold gas) as a function of baryon mass.
The low mass galaxies are gas rich, while the most massive systems
are predominantly stellar.  This trend is physically plausible,
but it is almost certainly exaggerated by our underestimate of
star formation rates in poorly resolved systems.  It should
therefore be treated as a tentative prediction, awaiting confirmation
with higher resolution simulations.

\begin{figure*}
\centerline{
\epsfxsize=4.5truein
\epsfbox[90 415 470 725]{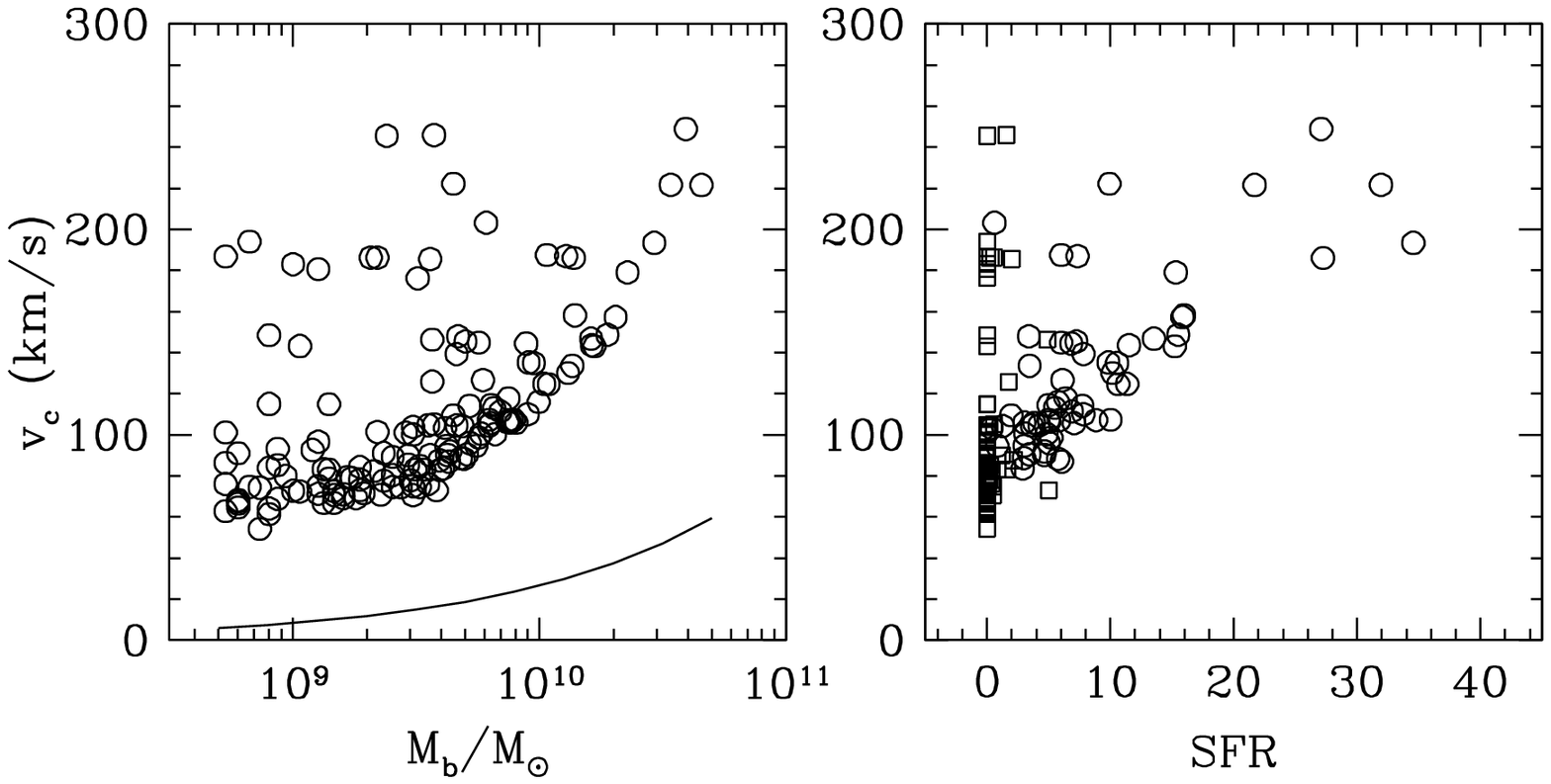}
}
\caption{
\label{fig:vc40}
Circular velocity $v_c=(GM/r)^{1/2}$ computed based on the total mass
within $r=40\hkpc$ (physical) around simulated galaxies in the LCDM
model at $z=3$, as a function of
galaxy baryon mass (left) and instantaneous star formation rate (right).
The solid line in the left hand panel shows the contribution to the
circular velocity from condensed baryons alone, $v_c=(GM_b/r)^{1/2}$,
again for $r=40\hkpc$.  In the right hand panel, squares represent
galaxies with $M_b < 60m_{\rm SPH} = 4\times 10^9 M_\odot$.
}
\end{figure*}

All of the simulated galaxies reside in dark matter halos,
and the more massive halos frequently contain several galaxies
(see \citealt{gardner00}, figure 5).  As a characteristic
of these halos, we calculate the circular velocity
$v_c=(GM/r)^{1/2}$ from the total mass (dark plus baryonic)
within a physical radius $r=40\hkpc$ around each
simulated galaxy.  Figure~\ref{fig:vc40} plots these
circular velocities against the baryon masses (left) and 
star formation rates (right) of the LCDM galaxies at $z=3$.
There is a well defined lower ridge line in the $v_c$ vs. $M_b$
plot, but there are also galaxies with low $M_b$ and high $v_c$,
which are usually ``satellite'' galaxies in halos with several
distinct baryonic subunits.
The galaxies with high SFR tend to reside in relatively 
massive halos, but halos above a given $v_c$ threshold
host galaxies with a wide range of SFR, and the
correlation between SFR and halo $v_c$ becomes increasingly
weak below SFR$\;=10\msunyr$.  Changing the radius for $v_c$
definition from $r=40\hkpc$ to $r=20\hkpc$ or $r=80\hkpc$
makes little difference to the appearance of Figure~\ref{fig:vc40}.
For the galaxies with high SFR, the circular velocities
in Figure~\ref{fig:vc40} are large compared to the typical 
nebular emission line widths $\sigma \approx 70\;\kms$ 
measured in LBGs \citep{pettini98}, but it is unclear that 
emission line widths have much to do with the dark matter potential
well depth even in local starburst galaxies \citep{lehnert96}.

\begin{figure*}
\centerline{
\epsfxsize=4.2truein
\epsfbox[85 40 550 730]{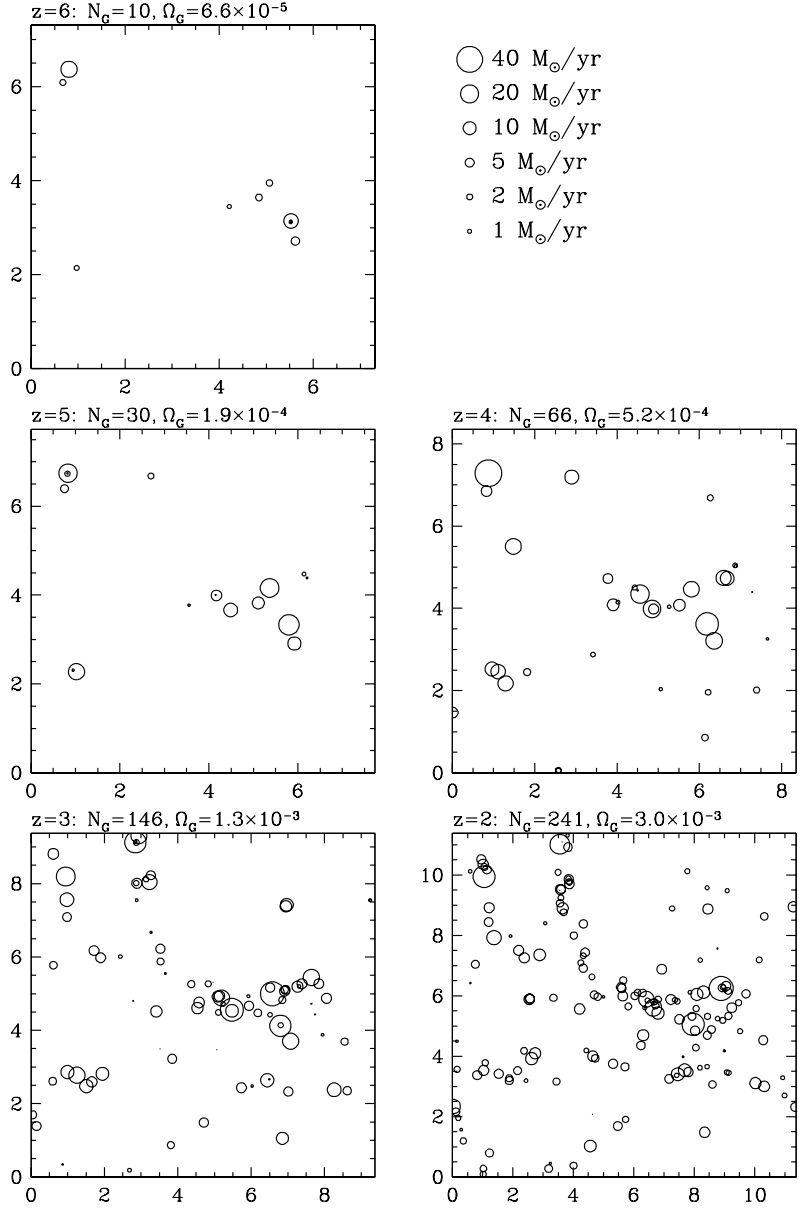}
}
\caption{
\label{fig:sfrCircle}
Galaxies in the LCDM model at $z=6$, 5, 4, 3, and 2.  
Each open circle represents a galaxy, with the area of the circle
proportional to the galaxy's instantaneous star formation rate.
The label above each panel indicates the redshift, the number
of galaxies, and the contribution to the density parameter
from material in galaxies (cold gas + stars).
Positions are marked in arc-minutes; the box size is
11.111 comoving $\hmpc$ at all redshifts.
}
\end{figure*}

Figure~\ref{fig:sfrCircle} illustrates the build-up of the 
galaxy population in the LCDM model from $z=6$ to $z=2$.
Each panel shows a projection of the box at the indicated
redshift, with each galaxy represented by a circle whose
area is proportional to the instantaneous star formation rate.
By $z=6$ there are already ten galaxies in the box
with SFR$\;>1\msunyr$ and two with SFR$\;>10\msunyr$.
As time goes on, the star formation rates of the most
massive galaxies tends to increase, though this trend saturates
after $z=4$.  The total number of star-forming galaxies in
the box increases steadily, with more than 200 galaxies
in the box at $z=2$.  Most strikingly, the locations of newly
forming galaxies are strongly correlated with the locations
of pre-existing galaxies, so that the backbone of structure
traced by the galaxy population remains similar from $z=6$ to $z=2$,
although it becomes better defined as more galaxies form.
We show in KHW that the galaxy correlation function $\xi(r)$
remains nearly constant in comoving coordinates from $z=4$
to $z=2$ and displays little dependence on the cosmological
model, although the amplitude of $\xi(r)$ does increase at a given
redshift if one selects only the most massive systems.  Comparisons of other
models in forms similar to Figure~\ref{fig:sfrCircle} appear
in KHW (figure 1).

\section{High-redshift Star Formation in CDM Cosmologies}
\label{sec:csf}

We now turn to the main quantitative results of the paper,
predictions of the star formation rates of high-redshift galaxies
in our simulations of the five CDM models listed in Table~\ref{tbl-1}.
Each simulation represents a theoretical model that
combines cosmological assumptions with assumptions about galactic
scale star formation.  While the predictions are not sensitive to 
the value of $c_*$, the one free parameter in our star formation
algorithm, they do depend on the general features
of the algorithm itself: the star formation rate is an increasing function
of the local density of cold gas,
and supernova feedback energy is deposited mainly
in the dense interstellar medium of forming galaxies and is therefore
radiated away rather efficiently.
We return to this point in \S\ref{sec:disc}.

\begin{figure*}
\centerline{
\epsfxsize=4.2truein
\epsfbox[65 55 535 715]{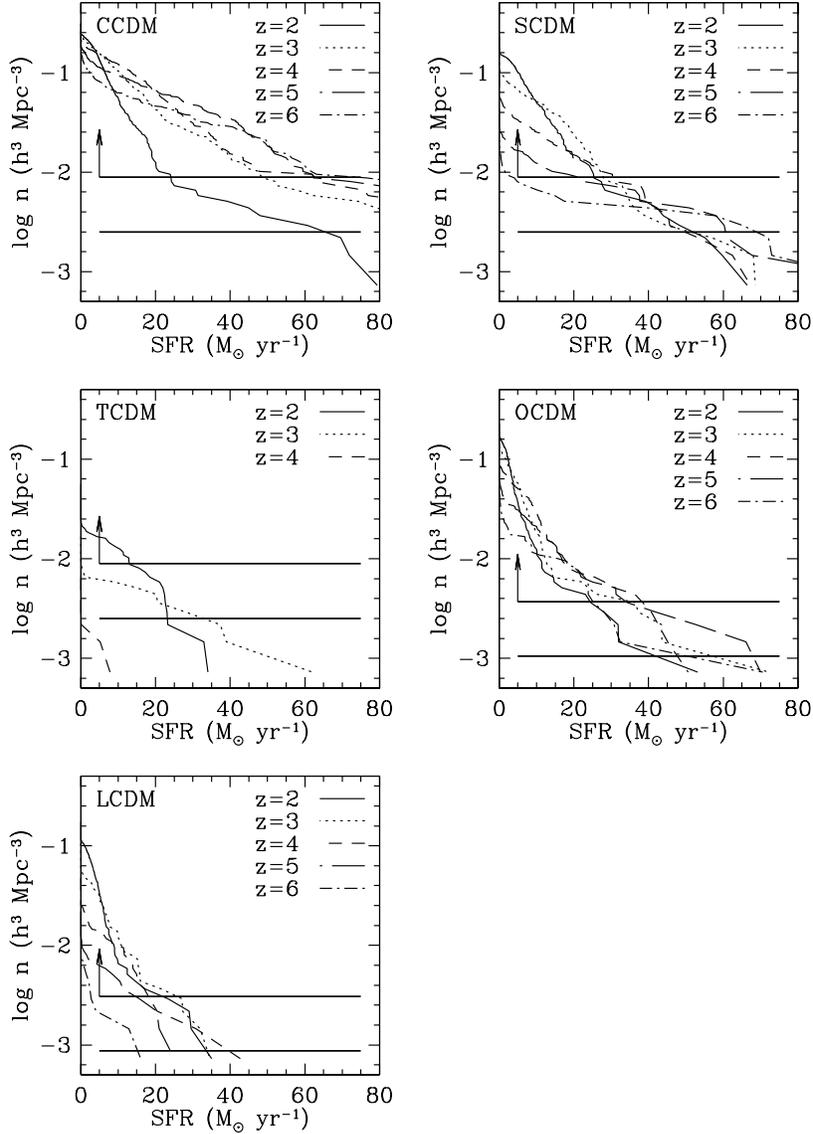}
}
\caption{
\label{fig:cumSfr}
The cumulative distribution function of simulated galaxies in terms
of instantaneous star formation rate (SFR).  Each panel represents
a different cosmological model, as indicated.  Horizontal lines
show the approximate comoving space density of spectroscopically 
confirmed, Lyman-break galaxies at $z=3$ in the samples of 
Steidel et al.\ (\citeyear{steidel96}; lower lines) 
and Lowenthal et al.\ (\citeyear{lowenthal97}; upper
lines, with an arrow indicating the possible effects of incompleteness
in the spectroscopy).
If these magnitude-limited samples select the objects
with the highest star formation rates, then the intersections of the
$z=3$ (dotted) curves with these horizontal lines indicate the predicted
star formation rates for objects at the limits of these samples.
}
\end{figure*}

Figure~\ref{fig:cumSfr} shows the cumulative distribution of
the simulated galaxies as a function of instantaneous SFR, at
redshifts $z=6$, 5, 4, 3, and 2.  The vertical axis represents the
comoving space density of all simulated galaxies with star formation
rate exceeding the indicated SFR, in $h^3\;{\rm Mpc}^{-3}$.
The amplitudes and redshift evolution of these distributions depend
strongly on the amplitude of mass fluctuations in the cosmological model
(see Figure~\ref{fig:pk}).
The CCDM simulation, with the highest fluctuation amplitude, already has
15 galaxies with SFR$\;>60\msunyr$ by $z=6$.  The number of
rapidly star-forming galaxies declines slowly from $z=5$ to $z=3$
and drops substantially between $z=3$ and $z=2$, though even at this
redshift the CCDM model has the highest overall star formation rate
of any of the models.  The TCDM model, with the lowest fluctuation 
amplitude, has no star formation in galaxies resolved by our simulation
until $z=4$.  It exhibits a rapid rise in the number of star-forming
galaxies between $z=4$ and $z=3$, and a steepening of the distribution
function from $z=3$ to $z=2$.  The flatness of the $z=3$ distribution
function may be in part a numerical artifact, since the lower mass
systems in this low amplitude model are barely resolved, and their
star formation rates may be correspondingly underestimated.
The other three models, with intermediate fluctuation amplitude,
show intermediate behavior.  For example, the LCDM model displays
a steady rise in the star formation rate from $z=6$ to $z=4$, then
little change from $z=4$ to $z=2$.

To facilitate comparison between Figure~\ref{fig:cumSfr} and
existing or future observational data, we list in Table~\ref{tbl-2}
the conversions from SFR to apparent magnitude and from 
comoving $h^3\;{\rm Mpc}^{-3}$ to number per arcmin$^2$
per unit redshift.  Specifically, $C_V$ is the volume conversion factor and
$m_{10}$ is the apparent magnitude
on the AB system at observed wavelength 
$\lambda_{\rm obs}=1500\times (1+z)$\AA\ that corresponds to a
star formation rate SFR$\;=10\msunyr$.  
The value of $m_{10}$ is similar in our high- and low-density models
because the effect of low-$\Om$ is approximately cancelled by
the increase in $H_0$.  We adopt the conversion
from SFR to UV continuum luminosity quoted by \citet{pettini98},
$L_{1500}=10^{29} 
({\rm SFR}/10\msunyr)\;{\rm erg}\;{\rm s}^{-1}\;{\rm Hz}^{-1}$,
which in turn is based on \citet{bruzual93} population synthesis
models assuming continuous star formation and a Salpeter initial
mass function extending from $0.1\msun$ to $100\msun$.
For example, the LCDM model predicts 
$10^{-2.66}$ galaxies per $h^{-3}\;{\rm Mpc}^3$ with SFR
above $20\msunyr$ at $z=5$, corresponding to 
$C_V \times n = 664 \times 10^{-2.66} = 1.46$ galaxies 
per arcmin$^2$ per unit redshift with apparent AB magnitude
less than $25.45 - 2.5\times\log(20/10) = 24.70$ at
$\lambda_{\rm obs}=9000$\AA.  The conversions in Table~\ref{tbl-2}
can be calculated using the formulas in \citet{hogg99} together
with the definition $m_{AB}=-2.5\log f_\nu - 48.60$.
Note, however, that these magnitudes assume no dust extinction,
while the UV continuum slopes of observed Lyman-break galaxies
imply typical UV extinctions of $0.5-2.5$ magnitudes 
\citep{adelberger00}.

The horizontal lines in Figure~\ref{fig:cumSfr} mark the comoving
space densities of spectroscopically confirmed objects in the LBG
samples of \citet{steidel96} and \citet{lowenthal97}, 
which have a mean redshift $z \approx 3$.  
Assuming that these surveys pick out the galaxies with the highest
star formation rates, the intersections of the dotted ($z=3$) simulation 
curves with these horizontal lines yield the predicted star formation
rates for galaxies near the sample magnitude limits.
The \citet{steidel96} magnitude limit of ${\cal R} \approx 25.5$
corresponds to a star formation rate of approximately $4 \msunyr$
with the assumptions described above (see Table~\ref{tbl-2}),
but $1-2.5$ magnitudes of dust extinction would increase the
implied SFR by factors of $2.5-10$.  The CCDM simulation predicts
a star formation rate of $\sim 90 \msunyr$ for objects with this
space density, which is clearly too high unless the true dust
extinction is much larger than \citet{adelberger00} estimate.
The other simulations predict star formation rates of 
$\sim 30-50\msunyr$, which is consistent with the \citet{steidel96}
results if the typical extinction is $\sim 2-2.5$ magnitudes.
The \citet{lowenthal97} survey of the HDF probes one magnitude deeper
than the \citet{steidel96} sample, and thus a factor of 2.5 lower in SFR.
The comoving space density of the \citet{lowenthal97} galaxies is
higher by a factor of 3.5, and since \citet{lowenthal97} only 
observed $\sim 2/3$ of their Lyman-break candidates (with a 44\%
success rate), the true space density at this magnitude limit
could be higher by a factor of $1.5-3$, as indicated by the 
arrow in Figure~\ref{fig:cumSfr}. 
The predicted star formation rates in the CCDM
simulation again appear too high, while the predictions of the
SCDM, OCDM, and LCDM models appear roughly compatible with the
\citet{lowenthal97} results if the dust extinction correction is
fairly large and the true space density is $2-3$ times the lower limit.
The TCDM model predicts very low star formation rates at the
\citet{lowenthal97} space density.  The numerical prediction is
clearly ruled out by the data, though this failure of the TCDM
model should still be viewed with some caution until it is confirmed
at higher numerical resolution.  We also note that the small 
size of the simulation volumes leads to significant uncertainties in
the predicted star formation rates at these low comoving densities:
the \cite{steidel96} space density corresponds to 3.5, 1.5, and 1.2
galaxies in the $(11.111 \hmpc)^3$ simulation volume
for the critical density, open, and flat-$\Lambda$ models, respectively.

\begin{figure*}
\centerline{
\epsfxsize=3.8truein
\epsfbox[90 105 470 725]{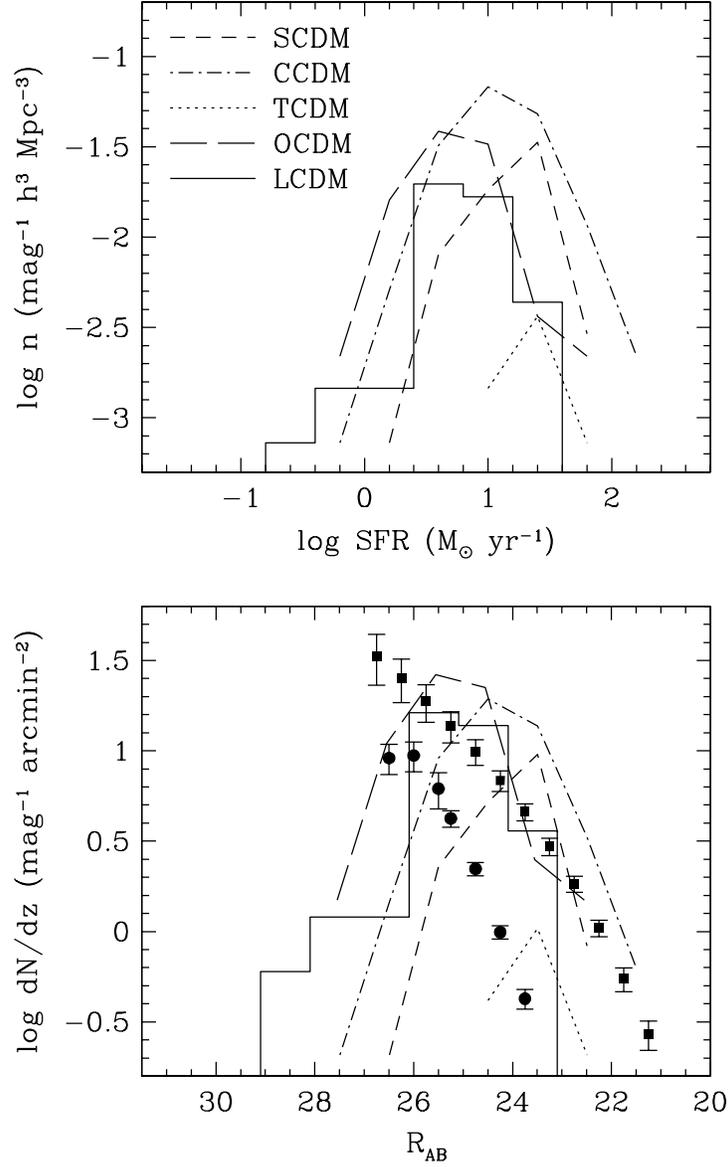}
}
\caption{
\label{fig:diffSfr}
{\it Top:} Differential distributions of instantaneous star formation rates
at $z=3$ in the five cosmological models.  For visual clarity, only
the LCDM result is shown as a histogram, but all distributions are
computed as histograms in 1-magnitude (0.4 dex) bins.  Only galaxies with 
baryonic mass $M>60m_{\rm SPH}$ are included.
{\it Bottom:} Same as top, but star formation rates have been translated
to corresponding $R_{\rm AB}$ magnitudes as described in the text,
and densities have been converted to directly observable units
of number per arcmin$^2$ per unit redshift.
Points show the luminosity function of Lyman-break galaxies at $z \sim 3$
estimated by 
Adelberger \& Steidel (\citeyear{adelberger00}, figure 11c) with (squares)
and without (circles) correction for dust extinction.
}
\end{figure*}

Figure~\ref{fig:diffSfr} presents a more detailed comparison between
numerical predictions and observational data at $z=3$.
The upper panel shows the differential distribution of star formation rates.
The LCDM result is shown as a solid histogram, but to preserve visual
clarity we show distributions for other models as connected lines.
Here we omit galaxies with baryonic mass $M_b<60m_{\rm SPH}$ because
limited numerical resolution would cause us to underestimate their
star formation rates.  The distributions therefore decline at low
SFR because of the absence of low mass galaxies.

In the lower panel we convert the predictions to observable units
using the conversions in Table~\ref{tbl-2}.  Points with $1\sigma$
statistical error bars show the luminosity function of
Lyman-break galaxies at $z \approx 3$ 
with (squares) and without (circles) correction for dust extinction,
from Adelberger \& Steidel
(\citeyear{adelberger00}; based on data from \citealt{steidel99}).
As \citet{adelberger00} emphasize, the extinction-corrected points
are quite uncertain: the corrections assume that the correlation
between UV continuum slope and extinction observed in local 
starburst galaxies \citep{meurer99} also holds at $z=3$, and the
points at faint magnitudes rely on an extrapolation of the luminosity
function.  However, these extinction corrections are probably the
best that can be made with current data, and they yield plausible
consistency between the population of UV-selected Lyman-break
galaxies and constraints on dust emission from sub-mm counts
and the infrared background \citep{adelberger00}.

Given the theoretical and observational uncertainties 
(which include numerical limitations, IMFs,
the value of $\Omb$, extinction corrections, incompleteness,
and $H_0$), we do not wish to draw strong 
conclusions from Figure~\ref{fig:diffSfr}.  The CCDM model appears
to predict excessively luminous galaxies, as expected from the
discrepancies already noted, and this discrepancy would be more
severe if the simulations had used the 
Burles \& Tytler (\citeyear{burles97}, \citeyear{burles98})
estimate of $\Omb$
instead of the \citet{walker91} estimate.
(Gardner et al.\ [in preparation] find that the SFR scales
approximately as $\Omega_b^{1.5}$.)
The galaxy population in the TCDM model is probably too faint,
unless the true extinction corrections are surprisingly small
or the numerical resolution effects are more severe than we think.
However, a higher $\Omb$ would improve the performance of this model.
The other three models appear roughly compatible with the data.
The limited dynamic range of the resolved galaxy populations in the
simulations prevents a detailed comparison to the shape of the
observed luminosity function.  In future work, we will combine
simulations of the LCDM model with different resolutions and box sizes
to represent the galaxy population over a wider mass range.

\begin{figure*}
\centerline{
\epsfxsize=4.5truein
\epsfbox[90 415 470 725]{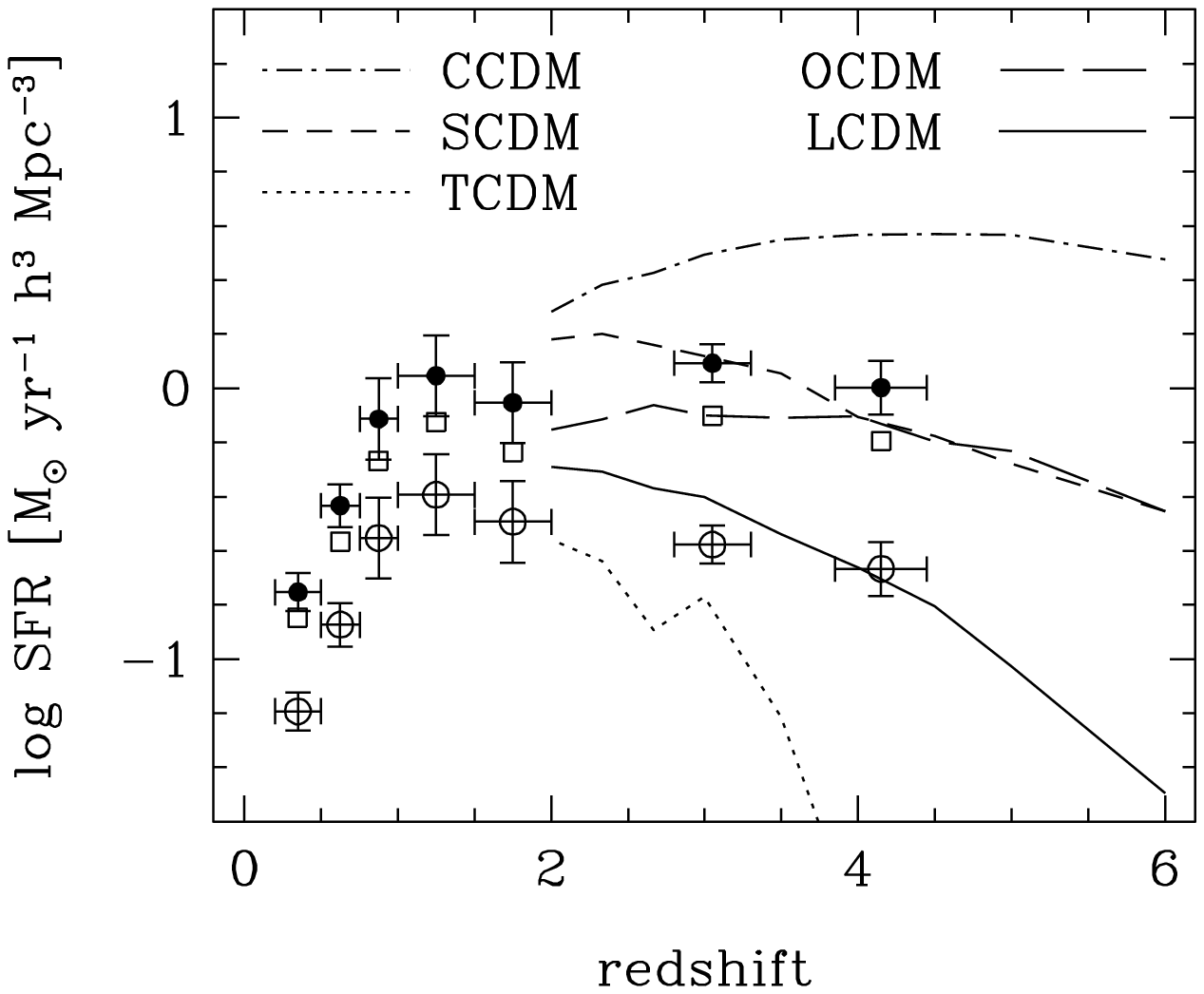}
}
\caption{
\label{fig:madau}
The average comoving density of star formation in the five cosmological
models, from $z=6$ to $z=2$.  Observational estimates with (filled circles)
and without (open circles) corrections for dust extinction are taken
from \citet{steidel99},
based on their own data ($z>3$) and
the data of Lilly et al.\ (\citeyear{lilly96}; $z<1$) 
and Connolly et al. (\citeyear{connolly97}; $1<z<2$);
they are computed for $\Om=1$, $\Ol=0$.
Open squares show the corresponding extinction-corrected estimates
for $\Om=0.4$, $\Ol=0.6$.
}
\end{figure*}

Figure~\ref{fig:madau} shows the globally averaged density of star formation
as a function of redshift, a representation of the cosmic star formation
history made famous by \citet{madau96}.  The simulation results accord
with the impressions from Figure~\ref{fig:cumSfr}.  In particular,
comparison of Figure~\ref{fig:madau} to Figure~\ref{fig:pk} shows that
the cosmic star formation history depends strongly on the amplitude
of mass fluctuations.  The high-amplitude, CCDM model predicts a 
high-amplitude SFR curve that peaks at high redshifts. 
The globally averaged SFR in this model declines slowly at $z<4$.  
In the other models,
the SFR appears to be reaching a plateau at $z\sim 2$, when the 
simulations stop.  The low-amplitude, TCDM model predicts the lowest SFR,
especially at high redshift.  
Data points in Figure~\ref{fig:madau} are taken from the 
analysis of \citet{steidel99}, based on their own data
and the data of \citet{lilly96} and \citet{connolly97}.
At each redshift, \citet{steidel99} estimate the global SFR
by integrating the UV continuum luminosity function for galaxies
with $L>0.1L_*$, so the points do not include the
contribution of the faintest galaxies (which are often below
the survey magnitude limits).
The open circles show estimates with no extinction correction,
while the filled circles incorporate extinction corrections
of 0.44 magnitudes at $z<2$ and 0.67 magnitudes at $z>2$.
Open squares show the extinction-corrected estimates converted
to the cosmological parameters of our LCDM model; points for
the OCDM model parameters would lie between the open squares
and filled circles.  

Unfortunately, the global SFR is a 
difficult quantity to predict robustly from numerical simulations 
with a limited dynamic range, since they miss the contribution from
galaxies below the resolution limit and underestimate the contribution 
from rare, massive galaxies, which are unlikely to form in a small
simulation volume.  Figure~2a of \cite{weinberg99} illustrates these
resolution and box size effects using several simulations of an LCDM 
model (one with a higher baryon density).
Because of these missing contributions, one should regard
the curves in Figure~\ref{fig:madau} 
as lower limits to the true predictions of these models.
The theoretical and observational uncertainties
make it difficult to draw strong conclusions from Figure~\ref{fig:madau}.
The CCDM model appears to predict too much star formation.
The SCDM and OCDM predictions agree fairly well with the
extinction-corrected estimates of \cite{steidel99}, but contributions
from lower mass galaxies or an increase in $\Omb$ would make this
agreement worse.  The LCDM predictions are somewhat low, but they might
plausibly be boosted towards reasonable agreement with higher 
resolution and/or higher $\Omb$.  The TCDM predictions are far
below the observational estimates.  Of course, most of the action
in the observational data takes place at $z<2$, after these
simulations stop.  Figure~2b of \citet{weinberg99} shows preliminary
results from simulations of a similar set of models
(from \citealt{dave99}), extended to $z=0$.  The global SFR
declines in all of the models at $z<1$, though not as sharply
as the data points in Figure~\ref{fig:madau}.

In Table~\ref{tbl-3}
we list a more robust prediction of the simulations, the contribution
to the globally averaged SFR from galaxies 
above our estimated resolution limit, those with $M_b>60m_{\rm SPH}$.
We also list the number density $n_{\rm res}$ of such galaxies at 
each redshift.  To the extent that galaxy absolute-magnitude correlates
with baryonic mass, the corresponding observational quantity could be
computed from a galaxy survey by choosing a limiting magnitude at
which the galaxy number density is $n_{\rm res}$ and summing the
contribution to the global SFR from galaxies above this magnitude limit.

\section{Discussion}
\label{sec:disc}

Our main result is that inflationary CDM models, combined with 
straightforward assumptions about galactic scale star formation,
predict a substantial population of star-forming galaxies at $z \geq 2$.
The stellar masses and star formation rates of these high-redshift
systems are sensitive to the amplitude of the underlying mass
power spectrum (compare, e.g., Figure~\ref{fig:pk} and Figure~\ref{fig:madau}.)
The results of the LCDM, OCDM, and SCDM simulations appear roughly
consistent with the observed properties of Lyman-break galaxies,
given the theoretical and observational uncertainties.  The 
low-amplitude, TCDM model predicts an anemic LBG population that
is probably inconsistent with current observations, though this
conclusion may be sensitive to our finite numerical resolution
and our adopted value of $\Omb$.  The high-amplitude, CCDM 
model appears to predict too much high-redshift star formation,
by a significant factor.

The \lya forest offers a more direct probe of the amplitude 
of mass fluctuations at high redshift \citep{croft98}.
Recent observational analyses 
(\citealt{croft99,mcdonald00}; Croft et al., in preparation) imply
that the matter power spectrum at $z \approx 2-3$ is similar to
that in our LCDM, OCDM, and SCDM models but incompatible with 
the CCDM or TCDM models.  It is reassuring that the models
supported by the \lya forest data appear to be the most compatible
with the star formation properties of observed LBGs, though an
increase in $\Omb$ to the values supported by recent D/H studies
(\citealt{burles97}, \citeyear{burles98}) and other \lya forest analyses
\citep{rauch97,weinberg97b} might spoil this agreement to some extent.
We will examine the influence of $\Omega_b$ on the high-redshift
galaxy population elsewhere (Gardner et al., in preparation);
our initial results imply that galaxy star formation rates in the 
SCDM model scale roughly as $\Omega_b^{1.5}$.
In KHW, we showed that the clustering of high-redshift
galaxies in these simulations is consistent with observed
LBG clustering \citep{adelberger98,giavalisco98}, and that
the clustering is insensitive to the cosmological model because
galaxies form at the same ``biased'' locations of the dark matter
distribution in all five simulations.

In \citet{gardner00}, we examine the predictions of these simulations for
damped \lya absorption, which is the other main observational 
probe of the high-redshift galaxy population.  The galaxies resolved
in these simulations account for only a fraction of the observed
damped \lya absorption at $z \approx 3$, ranging from $\sim 3\%$
in TCDM to $\sim 30\%$ in LCDM, SCDM, and CCDM to $\sim 50\%$ in OCDM.
Since the simulations already go to higher space densities than
existing spectroscopic samples of LBGs, our results imply that 
these LBG samples are not yet deep enough to include the galaxies
responsible for most damped \lya absorption.  \citet{haehnelt00}
reach a similar conclusion using analytic arguments.
By extrapolating the simulation results with the
aid of the Press-Schechter (\citeyear{press74}) mass function,
\cite{gardner00} 
conclude that absorption in lower mass systems is sufficient
to account for observed damped \lya absorption in any of these
cosmological models, with the possible exception of TCDM.
However, the uncertainties in the extrapolation are large,
and definitive examination of the compatibility between LBG
and damped \lya constraints will require higher resolution simulations.

Semi-analytic methods, sometimes combined with N-body simulations of
the dark matter distribution, are the main alternative to hydrodynamic
simulations for theoretical investigation of high-redshift galaxy formation.
Using these methods, several groups have found that CDM models like the
ones studied here can reproduce the numbers, luminosities, colors,
and clustering properties of observed LBGs 
\citep[e.g., ][]{baugh98,governato98,kauffmann99,somerville00}.
The semi-analytic papers have led to three rather different suggestions
about the nature of Lyman-break galaxies.  In the first view, observed
LBGs are the most massive galactic systems present at high redshift,
forming stars at a fairly steady rate \citep{baugh98}.  In the second
view, interactions play a crucial role in triggering bursts of star 
formation, and many LBGs are low mass systems boosted temporarily,
and briefly, to prominence \citep{kolatt99,somerville00}.  
A third, intermediate case is one in which LBGs are massive galaxies 
experiencing bursts of star formation stimulated by minor or major mergers
\citep{somerville00}.  This variety of views is mirrored to some extent 
in the observational literature on LBGs (compare, for example, 
\citealt{steidel96} to \citealt{lowenthal97} or \citealt{trager97}).

Our simulations suggest a picture that is intermediate between the
extremes of this debate, but closest to the first point of view.  
Star formation in the simulated galaxies is steady on timescales of 200 Myr
(Figure~\ref{fig:sfrComp}), and the instantaneous star formation 
rate is fairly well correlated with stellar mass (Figure~\ref{fig:sfrVsMstar}).
However, there is enough scatter in galaxy star formation rates
that a sample of galaxies selected above a SFR threshold includes
objects with a substantial range of stellar masses 
(Figure~\ref{fig:sfrVsMstar}), and these may reside in halos with a
wide range of circular velocities (Figure~\ref{fig:vc40}).
The simulations automatically include interactions and mergers, but they do not
resolve the existence of the low mass
systems envisioned to play an important role in the extreme version
of the collision-induced starburst model.

The properties of the simulated LBG population depend on the cosmological
initial conditions and on the basic physics of gravity and gas dynamics,
but they also depend on our adopted prescription for galactic scale
star formation.  The crucial features of this prescription are
(1) that the local star formation timescale decreases with increasing
gas density, as implied by studies of local galaxies \citep{kennicutt98}, 
and (2) that supernova feedback energy is deposited mainly in the dense
interstellar medium, where it is usually radiated away before it has
a large dynamical impact.  Since we do not require any external
triggers for star formation, an isolated galaxy that steadily 
accretes cold gas will convert that gas into stars, albeit over many
orbital times.  Interactions and mergers can enhance star formation
by driving gas to higher density, but galaxies do not build up 
large reservoirs of dense gas that wait to be ignited.
Limited resolution may reduce the influence of interactions and mergers
in these simulations, since they do not resolve low mass satellites
and do not resolve the nuclear star formation that is prominent
in high resolution simulations of starbursts 
induced by minor (\citealt{mihos94a}; \citeyear{mihos96})
or major \citep{mihos94b,hernquist95} mergers.  A more efficient feedback 
mechanism could also lead to more episodic star formation histories,
by producing cycles of starbursts followed by suppressed accretion
and cooling.  Since we have not investigated scenarios in which
interactions or feedback play a larger role, we cannot draw conclusions
about their viability.  However, our results suggest that the straightforward
treatment of star formation described in \S\ref{sec:numerics} is
sufficient to explain at least the basic properties of the observed 
LBG population within the CDM cosmological framework.

The clearest prediction of the simulations is that the Lyman-break
galaxies studied by 
Steidel et al.\ (\citeyear{steidel96}, \citeyear{steidel99}) and
\citet{lowenthal97} represent the tip of an iceberg.
The cumulative distribution curves in Figure~\ref{fig:cumSfr}
should be taken as lower limits to the predicted galaxy number
densities, especially at high redshifts, since limited resolution
causes these simulations to underestimate the star formation rates
in low mass systems, or to miss the systems entirely.
Nonetheless, the curves for, e.g., the LCDM model show that 
there should be large numbers of $z=3$ galaxies below the magnitude
limits of current LBG samples, and significant numbers of galaxies
with SFR$\;\geq 10 \msunyr$ even at $z \geq 5$.  Recent searches have
already yielded a number of spectroscopically confirmed galaxies
at $z=5-7$ \citep{spinrad98,weymann98,chen99,hu99}, 
and analyses of deep HST/NICMOS images 
show candidate objects to redshifts $z \geq 10$ \citep{yahata00,dickinson00}.
Systematic characterization of this faint galaxy population will be 
challenging, so it will be some time before simulations and data can be
compared quantitatively in the very high redshift regime.  But
the existence of a significant population of 
star-forming galaxies at $z > 5$ is a natural prediction of the CDM scenario.

Figures~\ref{fig:sfrVsMstar} and~\ref{fig:starFrac} also imply
some correlations between observable properties of $z=3$ galaxies.
While less massive galaxies tend to have lower star formation rates,
they usually have higher ratios of instantaneous SFR to time-averaged SFR,
and they should therefore have bluer spectral energy distributions
unless they are more heavily reddened by dust.
Less massive galaxies also tend to be more gas rich.
Unfortunately, both of these predicted trends could be exaggerated
by numerical resolution effects, so we do not regard them as very robust.

This paper presents a first attempt to characterize the star formation
properties of high-redshift galaxies using hydrodynamic simulations, 
but there is much room for progress with future simulations.
The emerging consensus on cosmological parameters, if it survives
the tightening of observational constraints, makes the task easier
by focusing effort on a preferred background model.  Within such
a framework, simulations with different box sizes and resolutions can be 
combined to model the galaxy population over a wider dynamic range of
mass and redshift, improving the comparison to the observed 
luminosity function (as in Figure~\ref{fig:diffSfr}) and global 
star formation history (as in Figure~\ref{fig:madau}).
\citet{weinberg99} take a first step along this path, using
multiple simulations of the LCDM model (with higher $\Omega_b$) to 
predict cumulative SFR distributions from $z=0.5$ to $z=10$.
Since the present simulations resolve galaxies far below the 
limits of current LBG spectroscopic samples, simulations of larger
volumes at lower resolution will improve the comparison between
predicted and observed LBG clustering.  Higher resolution simulations,
on the other hand, can probe the connection between LBGs and damped
\lya systems and test the robustness of some of the trends found
in \S\ref{sec:gals}.  They can also provide predictions of the structural
properties of high-redshift galaxies, such as size and morphology,
though these may be best investigated with simulations that zoom 
in to follow the formation of individual objects
\citep[e.g., ][]{haehnelt98,contardo98}.  
In the long run, we also hope to examine different formulations
of star formation and feedback, to determine what descriptions of
these physical processes match the observed properties of galaxies
and the intergalactic medium over the full range of accessible redshifts.

Where are the Lyman-break galaxies today?  Because our present simulations
stop at $z=2$, we will save a detailed examination of this question for
a future paper on the assembly history of galaxies, using simulations
(like those of \citealt{dave99}) that continue to $z=0$.
A first look at these simulations suggests that there is no simple,
one-sentence answer.  Between $z=3$ and $z=0$, galaxies accrete fresh
material and merge with each other, and many new galaxies form that
had no $z=3$ progenitors at all (at least above the numerical
resolution limits).  The particles that lie in galaxies at $z=3$ 
are distributed at $z=0$ among galaxies with a wide range of environments
and masses, though the most massive $z=0$ galaxies always contain some of 
these particles and the least massive often do not.  Any link between
LBGs and present-day ellipticals, or bulges, or halos, or clusters,
can at best capture a general trend, one that is likely to be violated
nearly as often as it is obeyed.  Fortunately, cosmological simulations
are an ideal tool for characterizing the full range of possible histories,
providing the theoretical thread that can tie snapshots of the galaxy
population at different redshifts into a coherent picture of galaxy
formation and evolution.

\acknowledgments

We thank Kurt Adelberger and Chuck Steidel for helpful discussions
and for providing the data plotted in Figures~\ref{fig:diffSfr}
and~\ref{fig:madau}.  We also thank James Lowenthal for helpful discussions.
This work was supported
by NASA Astrophysical Theory Grants NAG5-3922, NAG5-3820, and NAG5-3111,
by NASA Long-Term Space Astrophysics Grant NAG5-3525, and by the NSF under
grants ASC93-18185, ACI96-19019, and AST-9802568.  
The simulations were performed at the San Diego Supercomputer Center.

\vfill\eject

\vfill\eject

\begin{table}
\caption{Cosmological Model Parameters} \label{tbl-1}
\begin{center}\scriptsize
\begin{tabular}{crrrrrrrrrrr}
Model & $\Om$ & $\Ol$ & $h$ & $\Omb$ & $\sigma_{8}$ & $n$ 
& $60m_{\rm dark}$ & $60m_{\rm SPH}$\\
\tableline
CCDM  &     1.0  &    0.0  &     0.50  &    0.05  &    1.2   &   1.0 
      &     $1.7\times 10^{11}\msun$	&	$8.7\times 10^9\msun$ \\
SCDM  &     1.0  &    0.0  &     0.50  &    0.05  &    0.7   &   1.0 
      &     $1.7\times 10^{11}\msun$	&	$8.7\times 10^9\msun$ \\
OCDM  &     0.4  &    0.0  &     0.65  &    0.03  &    0.75  &   1.0 
      &     $5.0\times 10^{10}\msun$	&	$4.0\times 10^9\msun$ \\
LCDM  &     0.4  &    0.6  &     0.65  &    0.03  &    0.8   &   0.93 
      &     $5.0\times 10^{10}\msun$	&	$4.0\times 10^9\msun$ \\
TCDM  &     1.0  &    0.0  &     0.50  &    0.05  &    0.54  &   0.80
      &     $1.7\times 10^{11}\msun$	&	$8.7\times 10^9\msun$ \\

\end{tabular}
\end{center}
\end{table}

\begin{table}
\begin{tabular}{lllrrrrrrrrrr}
 \tableline\tableline
 \multicolumn{1}{c}{}& \multicolumn{1}{c}{}& 
 \multicolumn{1}{c}{ }& \multicolumn{2}{c}{$z=6$}& 
 \multicolumn{2}{c}{$z=5$}&  \multicolumn{2}{c}{$z=4$} &
 \multicolumn{2}{c}{$z=3$}&  \multicolumn{2}{c}{$z=2$} \\
\multicolumn{1}{l}{$\Om$}& 
\multicolumn{1}{l}{$\Ol$}& 
\multicolumn{1}{c}{$h$}& 
\multicolumn{1}{c}{$m_{10}$}& \multicolumn{1}{c}{$C_V$}&
\multicolumn{1}{c}{$m_{10}$}& \multicolumn{1}{c}{$C_V$}&
\multicolumn{1}{c}{$m_{10}$}& \multicolumn{1}{c}{$C_V$}&
\multicolumn{1}{c}{$m_{10}$}& \multicolumn{1}{c}{$C_V$}&
\multicolumn{1}{c}{$m_{10}$}& \multicolumn{1}{c}{$C_V$} \\ \tableline
1.0& 0.0& 0.50& 25.58& 191& 25.30& 217& 24.96& 249& 24.50& 285& 23.82& 314\\
0.4& 0.0& 0.65& 25.90& 621& 25.55& 654& 25.12& 680& 24.55& 684& 23.73& 631\\
0.4& 0.6& 0.65& 25.74& 590& 25.45& 664& 25.08& 746& 24.59& 824& 23.86& 847\\
\tableline\tableline
\end{tabular}
\caption{SFR and volume conversions: $m_{10}$ is the AB magnitude at observed
wavelength $\lambda_{\rm obs}=1500\times (1+z)$\AA\ for an object with
unobscured SFR$\;=10\msunyr$, and multiplication by $C_V$ converts space
densities from $h^3 \;{\rm Mpc}^{-3}$ to number per arcmin$^2$ per unit
redshift.}
\label{tbl-2}
\end{table}

\begin{table}
\begin{tabular}{lrrrrrrrrrr}
 \tableline\tableline
 \multicolumn{1}{c}{ }&
 \multicolumn{2}{c}{$z=6$}& 
 \multicolumn{2}{c}{$z=5$}&  \multicolumn{2}{c}{$z=4$} &
 \multicolumn{2}{c}{$z=3$}&  \multicolumn{2}{c}{$z=2$} \\
\multicolumn{1}{l}{Model}& 
\multicolumn{1}{c}{$n_{\rm res}$}& 
\multicolumn{1}{c}{SFR}&
\multicolumn{1}{c}{$n_{\rm res}$}& 
\multicolumn{1}{c}{SFR}&
\multicolumn{1}{c}{$n_{\rm res}$}& 
\multicolumn{1}{c}{SFR}&
\multicolumn{1}{c}{$n_{\rm res}$}& 
\multicolumn{1}{c}{SFR}&
\multicolumn{1}{c}{$n_{\rm res}$}& 
\multicolumn{1}{c}{SFR}\\
\tableline
CCDM&0.067& 2.733& 0.099& 3.386& 0.138& 3.380& 0.168& 2.909& 0.195& 1.748\\ 
SCDM&0.008& 0.341& 0.017& 0.515& 0.031& 0.708& 0.063& 1.204& 0.122& 1.411\\ 
TCDM&0.000& 0.000& 0.000& 0.000& 0.001& 0.010& 0.006& 0.168& 0.017& 0.268\\ 
OCDM&0.017& 0.326& 0.035& 0.554& 0.061& 0.732& 0.095& 0.739& 0.137& 0.658\\ 
LCDM&0.002& 0.023& 0.007& 0.091& 0.017& 0.204& 0.044& 0.380& 0.098& 0.490\\
\tableline\tableline
\end{tabular}
\caption{Comoving star formation density of resolved galaxies.
At each redshift, $n_{\rm res}$ is the comoving number density of simulated
galaxies with baryonic mass $M_b>60m_{\rm SPH}$, in 
$h^3\;{\rm Mpc}^{-3}$, and SFR is the contribution of these galaxies
to the comoving star formation density, in $\msunyr h^3\;{\rm Mpc}^{-3}$.
}
\label{tbl-3}
\end{table}
  
\end{document}